\documentclass[%
 reprint,
 amsmath,amssymb,
 aps,
 prd,
]{revtex4-2}

\usepackage{graphicx}
\usepackage{dcolumn}
\usepackage{bm}
\usepackage{hyperref}

\usepackage{aas_macros}

\usepackage{color}
\definecolor{bbsalmon}{rgb}{1.0, 0.47, 0.42}

\begin{document}

\title{Fast Bayesian analysis of individual binaries in pulsar timing array data}

\author{Bence B\'ecsy}
\email{bencebecsy@montana.edu}
\author{Neil J.~Cornish}%
\author{Matthew C.~Digman}%
\affiliation{%
 eXtreme Gravity Institute, Department of Physics, Montana State University, Bozeman, Montana 59717, USA
}%


\date{\today}

\begin{abstract}
Searching for gravitational waves in pulsar timing array data is computationally intensive. The data is unevenly sampled, and the noise is heteroscedastic, necessitating the use of a time-domain likelihood function with attendant expensive matrix operations. The computational cost is exacerbated when searching for individual supermassive black hole binaries, which have a large parameter space due to the additional pulsar distance, phase offset and noise model parameters needed for each pulsar. We introduce a new formulation of the likelihood function which can be used to make the Bayesian analysis significantly faster. We divide the parameters into \emph{projection} and \emph{shape} parameters. We then accelerate the exploration of the projection parameters by more than four orders of magnitude by precomputing the expensive inner products for each set of shape parameters. The projection parameters include nuisance parameters such as the gravitational wave phase offset at each pulsar. In the new scheme, these troublesome nuisance parameters are efficiently marginalized over using multiple-try Markov chain Monte Carlo sampling as part of a Metropolis-within-Gibbs scheme. The acceleration provided by our method will become increasingly important as pulsar timing datasets rapidly grow. Our method also makes sophisticated analyses more tractable, such as searches for multiple binaries, or binaries with non-negligible eccentricities.
\end{abstract}

\maketitle


\section{Introduction}
\label{sec:intro}

Gravitational waves (GWs) with nHz frequencies can be probed by monitoring the time-of-arrival (TOA) of radio pulses emitted by millisecond pulsars. The primary targets of these pulsar timing arrays (PTAs) are GWs from supermassive black hole binaries (SMBHBs). These can potentially be recovered from the data individually, or one can detect a stochastic GW background (GWB) emerging from the ensemble of all SMBHBs in the observable universe (for a review, see e.g.~\cite{SteveBook}).

All three major PTA experiments recently detected a common red noise process: the North American Nanohertz Observatory for Gravitational Waves (NANOGrav \cite{nanograv_12p5yr_gwb}); the European Pulsar Timing Array (EPTA \cite{epta_dr2_gwb}); and the Parkes Pulsar Timing Array (PPTA \cite{ppta_dr2_gwb}). The International Pulsar Timing Array (IPTA) also found strong support for such a signal in the combined dataset from these regional PTAs \cite{ipta_dr2_gwb}. A common red noise process is expected to be the first sign of the GWB \cite{Romano_crn_vs_hd, Astro4Cast}. There remain alternative explanations, which can be ruled out if we observe the Hellings-Downs correlations characteristic of a GWB. Assuming the signal is due to the GWB, we expect to see a clear sign of these correlations within the next few years \cite{Astro4Cast}.

The detection of GWs from the brightest individual SMBHB is expected to happen not long after the confident detection of the GWB (see e.g.~\cite{nature_of_first_det, LukeSMBHCatalog}). Searching for GWs from individual SMBHBs is especially important since they will provide direct evidence of SMBHBs emitting GWs, unlike the GWB, which in principle could come from other GW sources (e.g.~cosmological phase transitions \cite{nanograv_12p5yr_cosmoPT}). An individually resolved SMBHB would also provide a unique opportunity for joint electromagnetic and GW observations (see e.g.~\cite{charisi_smbhb_mma}). Several searches for individual SMBHBs have been carried out in the past, resulting in upper limits on the amplitude of GWs from SMBHBs \cite{nanograv_11yr_cw, epta_dr1_cw, ppta_dr1_cw}.

Numerous analysis techniques have been developed to search for and characterize GWs from individual SMBHBs (see e.g.~\cite{NeilCWMethods, Lee_et_al_CW_methods, JustinCWMethods, Steve_accelerated_CW}). Fully Bayesian methods tend to be computationally intensive due to the large parameter space that needs to be explored, and because the unevenly sampled datasets and heteroscedastic (non-stationary) noise rule out the fast Fourier domain methods used in the analysis of ground-based GW detectors (see \cite{LIGO_data_guide} and references therein). This problem will further intensify as our datasets get larger, and as we try to incorporate more complicated signal models, like eccentric binaries \cite{PTA_eBBH_newtonian, PTA_eBBH_PN} or multiple binaries \cite{BayesHopper}. In this paper we present a method that can significantly speed up such Bayesian analyses by separating \emph{shape parameters}, which determine the morphology of the GW signal, from \emph{projection parameters}, which only affect how the signal is projected onto the line-of-sight of each pulsar. Our new technique enables exploration of projection parameters (most of which are nuisance parameters) at practically zero cost. Due to the large number of projection parameters, this results in a significant speedup of the overall analysis. This method is implemented in the \texttt{QuickCW} software package \footnote{Publicly available at:\url{https://github.com/bencebecsy/QuickCW}}. Note that this approach is similar to the F-statistic method introduced in Ref.~\cite{Ellis_Fstat}, but instead of analytically maximizing over certain parameters, we numerically marginalize over them, thus keeping the analysis fully Bayesian. There are also similarities with the techniques presented in Ref.~\cite{Steve_accelerated_CW}, where the likelihood function is either maximized or marginalized over the pulsar phase parameters.

The paper is organized as follows. In Section \ref{sec:fast_likelihood}, we review the traditional formulation of the individual source signal model and introduce the alternative formulation allowing for the rapid exploration of projection parameters. In Section \ref{sec:sampling}, we describe sampling methods that can be used to maximize the advantage brought about by the new likelihood. We validate these methods by analyzing various simulated datasets (Section \ref{sec:results}) and the NANOGrav 11-year dataset \cite{nanograv_11yr_data, nanograv_11yr_cw} (Section \ref{sec:11yr_comparison}). We conclude and outline possible future directions in Section \ref{sec:conclusion}. Throughout this paper we use units where $G=c=1$.

\section{Fast likelihood}
\label{sec:fast_likelihood}

In this section we review the effect of GWs from a circular SMBHB on PTA residuals, and we describe an alternative formulation of the signal model allowing for the separation of shape and projection parameters. The latter shows several similarities with the F-statistic (see e.g.~\cite{Fstat_LIGO, Schutz_Fstat, Fstat_LISA, Fstat_PTA, Ellis_Fstat}). However, we keep all parameters of the signal free, instead of maximizing over some of them as is done in the F-statistic analysis.

The emitted GW signal can be written as \footnote{Note: the following notation follows the sign convention of the \texttt{ENTERPRISE} software package, which is slightly different than the conventions usually used in the literature}:
\begin{equation}
	h_{ab}(t,\hat{\Omega}) = e_{ab}^{+}(\hat{\Omega}) \; h_+(t,\hat{\Omega}) + e_{ab}^{\times}(\hat{\Omega}) \; h_\times(t,\hat{\Omega}) \,,
\end{equation}
where $\hat{\Omega}$ is a unit vector from the GW source to the Solar System barycenter (SSB), 
$h_{+,\times}$ are the polarization amplitudes, 
and $e_{ab}^{+,\times}$ are the polarization tensors. 
The polarization tensors can be written in the SSB frame as:
\begin{eqnarray}
	e_{ab}^{+}(\hat{\Omega}) &=& \hat{m}_a \; \hat{m}_b - \hat{n}_a \; \hat{n}_b \,, \\
	e_{ab}^{\times}(\hat{\Omega}) &=& \hat{m}_a \; \hat{n}_b + \hat{n}_a \; \hat{m}_b \,,
\end{eqnarray}
where $\hat{\Omega}$, $\hat{m}$, and $\hat{n}$ are orthonormal vectors defined as:
\begin{eqnarray}
	\hat{\Omega} &=& -\sin\theta \cos\phi \; \hat{x} - \sin\theta \sin\phi \; \hat{y} - \cos\theta \; \hat{z} \,, \\
	\hat{m} &=& \sin\phi \; \hat{x} - \cos\phi \; \hat{y} \,, \\
	\hat{n} &=& -\cos\theta \cos\phi \; \hat{x} - \cos\theta \sin\phi \; \hat{y} + \sin\theta \; \hat{z} \,.
\end{eqnarray}
The response of a pulsar to the source is described by the antenna pattern functions $F^+$ and $F^\times$:
\begin{eqnarray}
	F^+(\hat{\Omega}) &=& \frac{1}{2} \frac{(\hat{m} \cdot \hat{p})^2 - (\hat{n} \cdot \hat{p})^2}{1+\hat{\Omega} \cdot \hat{p}} \,, \\
	F^\times(\hat{\Omega}) &=& \frac{(\hat{m} \cdot \hat{p}) (\hat{n} \cdot \hat{p})}{1+\hat{\Omega} \cdot \hat{p}} \,,
\end{eqnarray}
where $\hat{p}$ is a unit vector pointing from the SSB to the pulsar. The effect of a GW on a pulsar's TOAs can be written as:
\begin{equation}\label{signal}
	s(t, \hat{\Omega}) = F^+(\hat{\Omega}) \; \Delta s_+(t) + F^\times(\hat{\Omega}) \; \Delta s_\times(t) \,,
\end{equation}
where $\Delta s_{+,\times}$ is the difference between the signal induced at the pulsar and at the Earth 
(the so-called ``pulsar term'' and ``Earth term''), 
\begin{equation}
	\Delta s_{+,\times}(t) = s_{+,\times}(t_p) - s_{+,\times}(t) \,,
\end{equation}
where $t$ is the time measured at the SSB and $t_p$ is the corresponding time at pulsar $p$.
From geometry, we can relate $t$ and $t_p$ by:
\begin{equation}\label{pulsartime}
	t_p = t - L_p (1 + \hat{\Omega} \cdot \hat{p}) \,,
\end{equation}
where $L_p$ is the distance to the pulsar.

For a circular binary, at zeroth post-Newtonian (0-PN) order, $s_{+,\times}$ is given by:
\begin{eqnarray}
    s_+(t) &=& \frac{{\cal M}^{5/3}}{d_L \, \omega(t)^{1/3}} \left[ \sin 2\Phi(t) \, \left(1+\cos^2 \iota \right) \, \cos2\psi \right. \nonumber \\
			&& \left. + 2 \cos 2\Phi(t) \, \cos \iota \, \sin 2\psi \right] \,, \label{eq:signal1} \\
	s_\times(t) &=& \frac{{\cal M}^{5/3}}{d_L \, \omega(t)^{1/3}} \left[ -\sin 2\Phi(t) \, \left(1+\cos^2 \iota \right) \, \sin2\psi \right. \nonumber \\
			&& \left. + 2 \cos 2\Phi(t) \, \cos \iota \, \cos 2\psi \right] \,, \label{eq:signal2}
\end{eqnarray}
where $\iota$ is the inclination angle of the SMBHB, $\psi$ is the GW polarization angle, 
$d_L$ is the luminosity distance to the source, 
and ${\cal M} \equiv (m_1 m_2)^{3/5}/(m_1+m_2)^{1/5}$ 
is a combination of the black hole masses $m_1$ and $m_2$ 
called the ``chirp mass.'' 

The evolution of the frequency in the $\Phi(t)$ phase terms can be kept fully general, or Taylor expanded to leading order in the first time derivative of the orbital angular frequency, $\dot \omega$. The signal in pulsar $p_i$ can be rewritten as (cf.~eq.~(\ref{signal})):
\begin{equation}
	s_i(t, \hat{\Omega}) = \sum_{j=1}^{4} \sigma_{4(i-1)+j}(\hat{\Omega}, A_{e}, \iota, \psi, \Phi_0, \Phi_i) S^{4(i-1)+j}(t),
\end{equation}
where the filter functions, $S^{4(i-1)+j}(t)$, are defined as:
\begin{eqnarray}
S^{4(i-1)+1}(t) &=& \left[\omega_0/\omega(t)\right]^{1/3} \cos 2 \Phi(t) \, , \nonumber \\
S^{4(i-1)+2}(t) &=& \left[\omega_0/\omega(t)\right]^{1/3} \sin 2 \Phi(t) \, , \nonumber \\
S^{4(i-1)+3}(t) &=& \left[\omega_i/\omega(t_{p_i})\right]^{1/3} \cos 2 \Phi(t_{p_i}) \, , \nonumber \\
S^{4(i-1)+4}(t) &=& \left[\omega_i/\omega(t_{p_i})\right]^{1/3} \sin 2 \Phi(t_{p_i}) \, , \label{eq:filters}
\end{eqnarray}
where the prefactors in front of the trigonometric functions take into account the slight change in amplitude due to the frequency changing within the observational timespan. The reference frequency of the Earth term is denoted as $\omega_0 = \omega(t_0)$, while $\omega_i = \omega(t_0-L_{p_i} (1 + \hat{\Omega} \cdot \hat{p}_i))$ is the reference frequency of the pulsar term in pulsar $p_i$. The time-dependent angular frequency is given by:
\begin{equation}
\omega(t) = \omega_0 \left[ 1 - \frac{256}{5} {\cal M}^{5/3}  \omega_0^{8/3} (t-t_0) \right]^{-3/8}.
\label{eq:omega_t}
\end{equation}

The phases in eq.~(\ref{eq:filters}) are given by:
\begin{eqnarray}
\Phi(t) &=& \Phi_0 + \frac{1}{32} {\cal M}^{-5/3} \left[ \omega_0^{-5/3} - \omega(t)^{-5/3} \right] \, , \nonumber \\
\Phi(t_{p_i}) &=& \Phi_{i}+ \frac{1}{32} {\cal M}^{-5/3} \left[ \omega_{i}^{-5/3} - \omega(t_{p_i})^{-5/3} \right]\, .
\end{eqnarray}
Note that while the initial phases $\Phi_{i}$ and $\Phi_0$ are not independent parameters, we treat them as independent since the current uncertainties in the pulsar distances are so large that it is practically impossible to phase connect the Earth and pulsar terms. The first two filters are for the Earth term, and are the same for all pulsars (but they are sampled at different discrete times due to the different observing schedules for each pulsar).
The coefficients $\sigma_k$ are given by:
\begin{widetext}
\begin{eqnarray}
\sigma_{4(i-1)+1} &=& -A_{\rm e} \omega_0^{-1} \left[ \cos 2\Phi_0 (1+\cos^2 \iota) (\cos 2\psi F^+_i - \sin 2\psi F^\times_i) - 2 \sin 2\Phi_0 \cos\iota (\sin 2\psi F^+_i + \cos 2\psi F^\times_i)\right] \, ,  \nonumber \\
\sigma_{4(i-1)+2} &=& -A_{\rm e} \omega_0^{-1} \left[ \sin 2\Phi_0 (1+\cos^2 \iota) (\cos 2\psi F^+_i - \sin 2\psi F^\times_i) + 2 \cos 2\Phi_0 \cos\iota (\sin 2\psi F^+_i +  \cos 2\psi F^\times_i )\right] \, ,  \nonumber \\
\sigma_{4(i-1)+3} &=& A_i \omega_0^{-1} \left[ \cos 2\Phi_i (1+\cos^2 \iota) (\cos 2\psi F^+_i - \sin 2\psi F^\times_i) - 2 \sin 2\Phi_i \cos\iota (\sin 2\psi F^+_i + \cos 2\psi F^\times_i)\right] \, ,  \nonumber \\
\sigma_{4(i-1)+3} &=& A_i \omega_0^{-1} \left[ \sin 2\Phi_i (1+\cos^2 \iota) (\cos 2\psi F^+_i - \sin 2\psi F^\times_i) + 2 \cos 2\Phi_i \cos\iota (\sin 2\psi F^+_i +  \cos 2\psi F^\times_i )\right] \, ,
\end{eqnarray}
\end{widetext}
where $A_{\rm e}={\cal M}^{5/3} d_L^{-1} \omega_0^{2/3}$ and $A_i={\cal M}^{5/3} d_L^{-1} \omega_i^{2/3}$. Note that $A_i$ are not independent parameters, since they are uniquely determined by $A_{\rm e}$, $\omega_0$, and $\omega_i$. The log likelihood can be written as:
\begin{equation}
\log L = -\frac{1}{2} (\delta t - s|\delta t -s) -\frac{1}{2} \log \det (2 \pi C) \, ,
\end{equation}
where:
\begin{equation}\label{inner_product}
 (a|b) = a^T C^{-1} b\, .
\end{equation}

Here $C=N+TBT^T$, where $N$ is the white noise covariance matrix, $T$ is the design matrix for the timing model, red noise and jitter noise, and $B$ is the prior matrix for the hyperparameters of those (see e.g.~\cite{SteveBook}). Using the Woodbury matrix identity, we can express the inverse of $C$ as:
\begin{equation}\label{cmatrix}
 C^{-1} = (N+TBT^T)^{-1} = N^{-1} - N^{-1} T \Sigma^{-1} T^T N^{-1}\, ,
\end{equation}
where $\Sigma^{-1} = B^{-1} - T^T N^{-1} T$. Using the four filters we have:
\begin{eqnarray}
\log L &=& -\frac{1}{2} (\delta t|\delta t) -\frac{1}{2} \log \det (2 \pi C) \nonumber \\
&& + \sum_{k=1}^{4 {N_{\rm p}}} \sigma_k N^k - \frac{1}{2} \sum_{k=1}^{4 {N_{\rm p}}} \sum_{l=1}^{4 {N_{\rm p}}}\sigma_k \sigma_l M^{kl}\, ,
\label{eq:logL}
\end{eqnarray}
where $N^k = (\delta t | S^k)$, $M^{kl} = (S^k | S^l)$, and $N_{\rm p}$ is the number of pulsars in the array. Computing these inner products is the expensive step. The inner products have to be recomputed each time the noise model or the shape parameters of the signal (see Table \ref{tab:parameters}) are updated. Note that while the vast majority of the off-diagonal terms in $M^{kl}$ are zero, the per-pulsar quadratures and pulsar-Earth cross terms will not vanish since the data is un-evenly sampled and the orbital periods are generally not integer sub multiples of the observation time. The band-diagonal structure of the $M^{kl}$ mean that it can computed and stored in an array of size $10 N_{\rm p}$, rather than the naive $(4 N_{\rm p})^2$. Note that in the presence of correlated noise between different pulsars (e.g.~a stochastic GW background), inner products will include cross-terms between pulsars. Appendix \ref{sec:gwb_appendix} describes how the cross terms can be avoided, thereby maintaining the factorized form of the likelihood and the band-diagonal structure of  the $M^{kl}$ matrix.

\begin{table}[htbp]
\caption{\label{tab:parameters}%
List of shape and projection parameters and their numbers as a function of the number of pulsars ($N_{\rm p}$).
}
\begin{ruledtabular}
\begin{tabular}{c|c}
Shape parameters & Projection parameters\\
($4 + 3 \times N_{\rm p}$) & ($4 + N_{\rm p}$)\\
$\theta$, $\phi$, $f_{\rm GW}$, $\mathcal{M}$, $L_{p_i}$, $\gamma_{p_i}$, $A_{RN,p_i}$ & $\iota$, $A_{\rm e}$, $\Phi_0$, $\Psi$, $\Phi_i$\\
\end{tabular}
\end{ruledtabular}
\end{table}

For a fixed set of noise parameters and shape parameters, we can compute the likelihood for {\em any} set of projection parameters (see Table \ref{tab:parameters}) in essentially zero time. See details of how we can take full advantage of the speedup by an implementation relying on \texttt{Numba} in Appendix \ref{sec:implementation_appendix}. This increased speed allows us to fully marginalize over these projection parameters by performing a large number of MCMC updates of just these parameters.

If we are only interested in the Earth term, the sky location could also be marginalized over without needing to recompute the inner products. The pulsar terms ruin this separation of variables since the pulsar time $t_p$ from eq.~(\ref{pulsartime}) depends on the sky location. 

We can see from eq.~(\ref{eq:omega_t}) that the initial angular frequency at each pulsar is given by:
\begin{equation}\label{pf}
	\omega_i = \omega_0 \left( 1 + \frac{256}{5} {\cal M}^{5/3} \omega_0^{8/3} L_{p_i} (1 + \hat{\Omega} \cdot \hat{p}_i)  \right)^{-3/8} \, .
\end{equation}
Note that the Earth term and pulsar term can be quite different since:
\begin{eqnarray}
\frac{256}{5} {\cal M}^{5/3} \omega_0^{8/3} L_{p_i} &=& 10.16 \left(\frac{{\cal M}}{10^9 \, M_\odot} \right)^{5/3} \nonumber \\
&& \times \left(\frac{\omega_0}{2\pi / {\rm yr}} \right)^{8/3} \left(\frac{L_{p_i}}{1 \, {\rm kpc}}\right)  \, .
\label{eq:PT_freq_evolve}
\end{eqnarray}
On the other hand, the Earth term frequency only changes by a small amount during an ${\cal O}(10)$ year observing span so long as the systems are not very heavy or the current frequency is not too high since:
\begin{eqnarray}
\frac{256}{5} {\cal M}^{5/3} \omega_0^{8/3} t &=&  0.031 \left(\frac{{\cal M}}{10^9 \, M_\odot} \right)^{5/3} \nonumber \\
&& \times  \left(\frac{\omega_0}{2\pi /{\rm yr}} \right)^{8/3} \left(\frac{t}{10 \ {\rm yr}} \right) \, .
\label{eq:inband_freq_evolve}
\end{eqnarray}


\section{Metropolis-within-Gibbs sampling}
\label{sec:sampling}

With the new formulation of the likelihood in eq.~(\ref{eq:logL}) the computational burden is highly dependent on which parameters we try to update. Since calculating the likelihood at new projection parameter values is significantly faster than calculating it for a new set of shape parameters, we propose more updates in projection parameters. We achieve that by employing a Metropolis-within-Gibbs sampler (see \cite{metropolis-within-gibbs} and references therein), where the sampler completes a block of projection parameter updates (in this case typically $\mathcal{O}(10^3)$) before attempting a shape parameter update. The Metropolis-within-Gibbs sampler allows us to sample projection parameters extremely well with practically no additional cost.

In order to optimize the mixing in shape parameters, we also use a technique called Multiple Try MCMC (MTMCMC, see e.g.~\cite{MTMCMC_original,MTMCMC_review}). The idea of MTMCMC is that one can propose $N$ different points, select one of them based on some importance weights, and accept or reject the new sample based on an acceptance probability which depends on the likelihood at all $N$ proposed points. With $N=1$, one recovers the Metropolis-Hastings algorithm, while as $N \to \infty$, we draw independent samples from the posterior. Variants of MTMCMC where the trials can be drawn from different distributions or can be correlated were introduced in \cite{correlated_MTMCMC, gibbs_MTMCMC}.

We apply MTMCMC for the shape parameter updates in our Metropolis-within-Gibbs sampler as follows. Let us denote the parameters of the MCMC chain at the $i$th iteration as:
\begin{equation}
\bm{\theta}_i = \left\{ \bm{\theta}^{(\rm s)}_i, \bm{\theta}^{(\rm p)}_i \right\},
\end{equation}
where $\bm{\theta}^{(\rm s)}_i$ are the shape parameters and $\bm{\theta}^{(\rm p)}_i$ are the projection parameters at the $i$th iteration. We determine the next sample, $\bm{\theta}_{i+1}$, using the following algorithm:
\begin{enumerate}
 \item Draw a set of new shape parameters $\bm{\theta}^{(\rm s)}$ from the proposal $q^{(\rm s)} \Big(\bm{\theta}^{(\rm s)} \Big| \bm{\theta}^{(\rm s)}_i \Big)$.
 \item Draw $N$ different sets of projection parameters, $\bm{\theta}^{(\rm p)}_{(1)}, ..., \bm{\theta}^{(\rm p)}_{(k)}, ..., \bm{\theta}^{(\rm p)}_{(N)}$, from the proposals $q^{(\rm p)}_k \Big(\bm{\theta}^{(\rm p)}_{(k)} \Big| \bm{\theta}^{(\rm p)}_i \Big)$.
 \item Randomly select $\bm{\theta}_{(j)} = \left\{ \bm{\theta}^{(\rm s)}, \bm{\theta}^{(\rm p)}_{(j)} \right\}$, according to the probability mass function:
 \begin{equation}
  p(\bm{\theta}_{(j)}) = \frac{L(\bm{\theta}_{(j)})}{\sum_{k=1}^{N} L(\bm{\theta}_{(k)})}.
  \label{eq:mtmcmc_pmf}
 \end{equation}
 \item Form $N$ auxiliary samples:
 \begin{equation}
  \bm{\Theta}_{(k)} =
  \begin{cases}
   \left\{ \bm{\theta}^{(\rm s)}_i, \bm{\theta}^{(\rm p)}_{(k)} \right\} &\mbox{if } k \ne j\\
   \bm{\theta}_{i} &\mbox{if } k = j.
  \end{cases}
 \end{equation}
 \item Set $\bm{\theta}_{i+1} = \bm{\theta}_{(j)}$ with probability:
 \begin{equation}
  {\rm min} \left( 1, \frac{\sum_{k=1}^N L(\bm{\theta}_{(k)})}{\sum_{k=1}^N L(\bm{\Theta}_{(k)})} \right),
  \label{eq:mtmcmc_accept}
 \end{equation}
 otherwise, set $\bm{\theta}_{i+1} = \bm{\theta}_{i}$.
\end{enumerate}
Note that eqs.~(\ref{eq:mtmcmc_pmf}) and (\ref{eq:mtmcmc_accept}) only depend on the likelihood values and not on the proposal densities, because we only employ symmetric proposal distributions. We can see that if $N=1$, eq.~(\ref{eq:mtmcmc_accept}) reduces to the usual Metropolis-Hastings acceptance probability. Also note that since we are selecting the proposed projection parameters, $\bm{\theta}^{(\rm p)}_{(j)}$, according to eq.~(\ref{eq:mtmcmc_pmf}), as $N \to \infty$ we are drawing independent samples from the conditional likelihood at the new shape parameters, $L(\bm{\theta}^{(\rm p)}_{(j)} | \bm{\theta}^{(\rm s)})$. At this large $N$ limit, we can also interpret the acceptance probability in eq.~(\ref{eq:mtmcmc_accept}) as comparing the likelihood at the new and old shape parameters marginalized over the the projection parameters.

To ensure the new set of shape parameters is not rejected due to the lack of an appropriate set of projection parameters, we use a fairly large number of trials (typically $N=10,000$). A large number of projection parameter trials increases acceptance of shape parameter proposals with little additional cost due to the comparatively cheap evaluation of the likelihood at different projection parameters. For the shape parameter proposals, we use a mix of Fisher proposals, differential evolution proposals, and prior draws. At a given step we only update a specific set of parameters: (i) the four common parameters (sky location, chirp mass, frequency); (ii) pulsar distances; (iii) red noise parameters. For the projection parameter proposals, we always have one trial keep its original projection parameters. This gives a good chance of accepting the new shape parameters even if for some reason the large number of trials with perturbed projection parameters would land on low-likelihood places. For the rest, we draw each projection parameter independently using an optimal jump scale determined by the second derivative of the likelihood in that direction. If the optimal scale is larger than a threshold, we do a draw from the prior instead. Switching to prior draws ensures that when a parameter's value is not well determined we do a prior draw resulting in good exploration. Details about the implementation of the new likelihood function and the sampler can be found in Appendix \ref{sec:implementation_appendix}.

In general, MTMCMC algorithms provide better mixing at the cost of additional likelihood evaluations needed in eqs.~(\ref{eq:mtmcmc_pmf}) and (\ref{eq:mtmcmc_accept}). MTMCMC methods are particularly well suited to our new method, because the extremely cheap evaluations of the likelihood for different projection parameters give us the benefits at little additional cost.

\section{Results with simulated data}
\label{sec:results}

To test and illustrate the performance of the new methods described above, we analyzed several simulated datasets. To gauge expected runtimes realistically, all our datasets are made to resemble the latest publicly available dataset of the NANOGrav collaboration, the NANOGrav 12.5-year dataset \cite{nanograv_12p5yr_data}. They all contain the same 45 pulsars with the same timing solution and same observation properties (epochs, observing frequencies, TOA errors) as the real dataset. We simulated white and red noise in all pulsars according to the best-fit parameters found for the real observations. The typical runtime on a dataset of this size and complexity was a few days on an AMD Ryzen Threadripper 3970X 32-core processor.

In addition to white and red noise, we added three different signals to our dataset to test our analysis in different interesting scenarios: a slowly evolving signal which has comparable frequencies in the Earth term and the pulsar terms (see Dataset 1 in Table \ref{tab:inj_params} and Section \ref{sssec:slowly_evolving}); a rapidly evolving signal where the Earth and pulsar terms have significantly different frequencies (see Dataset 2 in Table \ref{tab:inj_params} and Section \ref{sssec:fast_evolving}); and a low-SNR marginally detectable signal at the most sensitive sky location and frequency, which is meant to represent the kind of signal we are most likely to detect first (see Dataset 3 in Table \ref{tab:inj_params} and Section \ref{sssec:marginal}). We also analyzed a dataset without any signals to test how the new pipeline can produce upper limits if no significant GW sources are found (see Section \ref{ssec:upper_limit}).

\begin{table*}[htbp]
\caption{\label{tab:inj_params}%
List of parameter values for simulated datasets. Red noise parameters and pulsar distances were set to the official values from the NANOGrav 12.5-year dataset \cite{nanograv_12p5yr_data}. The GW phase at each pulsar was determined from $\Phi_0$, the light travel time to each pulsar, and $f_{\rm GW}$.
}
\begin{ruledtabular}
\begin{tabular}{c|ccccccccc}
 & $\theta$ & $\iota$ & $\phi$ & $f_{\rm GW}$ & $A_{\rm e}$ & $\mathcal{M}$ & $\Phi_0$ & $\Psi$ & $d_L$\footnote{Note that $d_L$ is not an independent paramter we fit for. It is completely determined by $A_{\rm e}$, $f_{\rm GW}$, and $\mathcal{M}$.}\\
\colrule
Dataset 1 \phantom{\Large{A}} & $\pi$/3 & 1.0 & 4.5 & 8 nHz & $5\times10^{-15}$ & $5\times10^{8}\ M_{\odot}$ & 1.0 & 1.0 & 7.5 Mpc\\
Dataset 2 \phantom{\Large{A}} & $\pi$/3 & 1.0 & 4.5 & 20 nHz & $1\times10^{-14}$ & $5\times10^{9}\ M_{\odot}$ & 1.0 & 1.0 & 320 Mpc\\
Dataset 3 \phantom{\Large{A}} & 2$\pi$/3 & 0.5 & 4.5 & 8 nHz & $2\times10^{-15}$ & $1\times10^{9}\ M_{\odot}$ & 2.0 & 1.5 & 60 Mpc\\
\end{tabular}
\end{ruledtabular}
\end{table*}

\subsection{Detection analyses}
\label{ssec:detection_runs}

\subsubsection{Slowly evolving signal}
\label{sssec:slowly_evolving}

\begin{figure}[htbp]
\includegraphics[width=0.92\columnwidth]{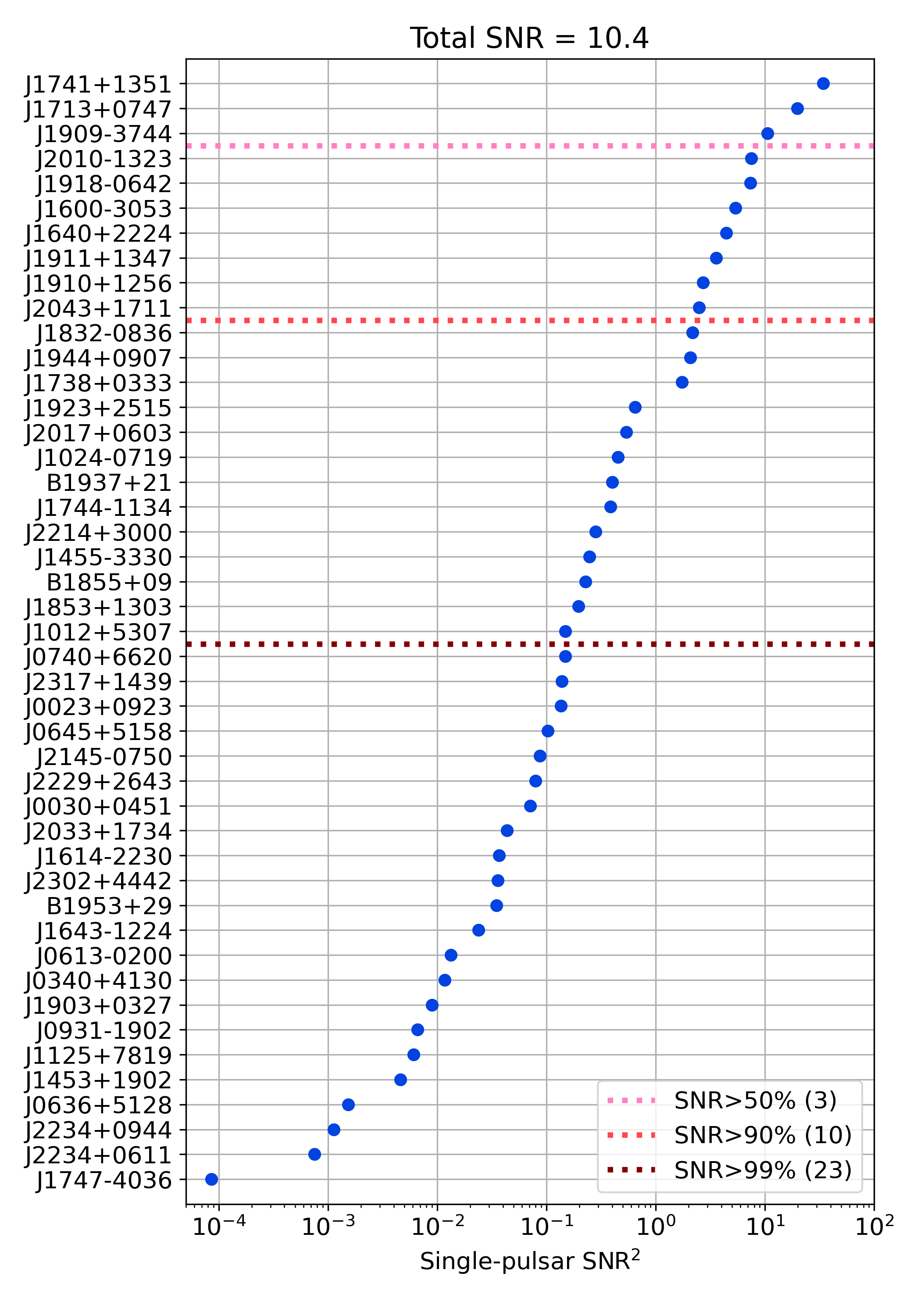}
\caption{SNR$^2$ values in datastreams of each simulated pulsar for slowly a evolving signal (Dataset 1 in Table \ref{tab:inj_params}). Horizontal dashed lines indicate the minumum number of pulsars needed to reach 50/90/99\% of the total SNR$^2$.}
\label{fig:snrs_5e-15}
\end{figure}

We first analyzed a signal with a relatively low chirp mass and frequency (see Dataset 1 in Table \ref{tab:inj_params}) resulting in only slightly different Earth term and pulsar term frequencies (cf.~eq.(\ref{eq:PT_freq_evolve})). We chose the amplitude of the signal to achieve a moderate SNR of 10.4. Fig.~\ref{fig:snrs_5e-15} shows how the square of the SNR (a good proxy for detectability) is distributed among the 45 pulsars in the array. The highly heterogeneous distribution means that only 3/10/23 pulsars are responsible for 50/90/99\% of the total SNR$^2$. There are two reasons for this: (i) the pulsars in the array have a wide range of observation timespans and timing precisions, since they correspond to the real pulsars in the NANOGrav 12.5-year dataset; (ii) pulsars in favorable sky locations can incur significantly higher SNRs compared to pulsars in ``bad'' sky locations. This imbalance in SNR might suggest that one can arrive at similar results by only using a small subset of the pulsars. However, even pulsars with negligible SNRs can contribute to parameter estimation, since they can rule out some parts of the parameter space where they would be able to reach a higher SNR. Also, one cannot know a priori which pulsars could be neglected, since even a less precisely timed pulsar can have a relatively high SNR if it happens to be in a favorable sky location for a particular source.

\begin{figure*}[htbp]
\includegraphics[width=0.95\textwidth]{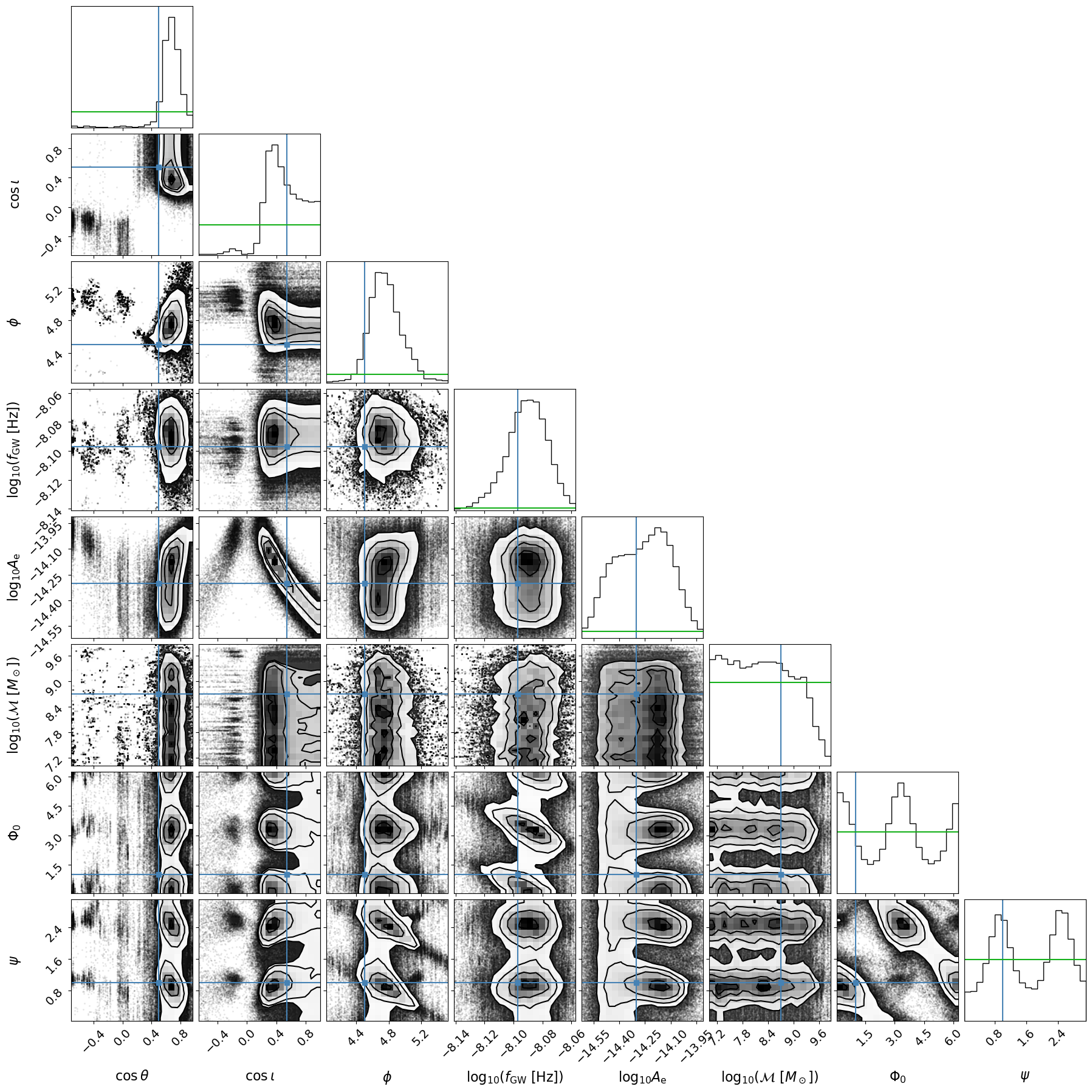}
\caption{Corner plot of parameters common to all pulsars for the slowly evolving signal (SNR=10.4, Dataset 1 in Table \ref{tab:inj_params}). We show the posterior with black, the prior with green and the true values of parameters with blue. Contour lines represent the 1/2/3-$\sigma$ levels in two dimensions, which correspond to 39.3\%/67.5\%/86.5\% credible regions.}
\label{fig:common_par_corner_5e-15_slow_evolve}
\end{figure*}

Fig.~\ref{fig:common_par_corner_5e-15_slow_evolve} shows the one-dimensional and two-dimensional marginal distributions of the 8 signal parameters common to all pulsars. Blue lines and dots indicate the true values of parameters, while green horizontal lines show the prior distribution for each parameter. The true values of parameters lie within the bulk of the posterior for all parameters. As expected given the slow frequency evolution of the signal, the chirp mass distribution is largely unconstrained, with only the largest values being ruled out as they would have resulted in a detectable frequency evolution. The amplitude is highly correlated with a number of nuisance parameters, most notably the inclination angle ($\iota$), emphasizing the importance of effectively sampling these. Some parameters show a highly complex, multimodal posterior, which makes the sampling of those parameters particularly challenging. Some of the intricate structure (especially in sky location) is due to the uneven distribution of pulsars on the sky, which results in highly variable sensitivity over the sky.



\begin{figure*}[htbp]
\includegraphics[width=0.98\textwidth]{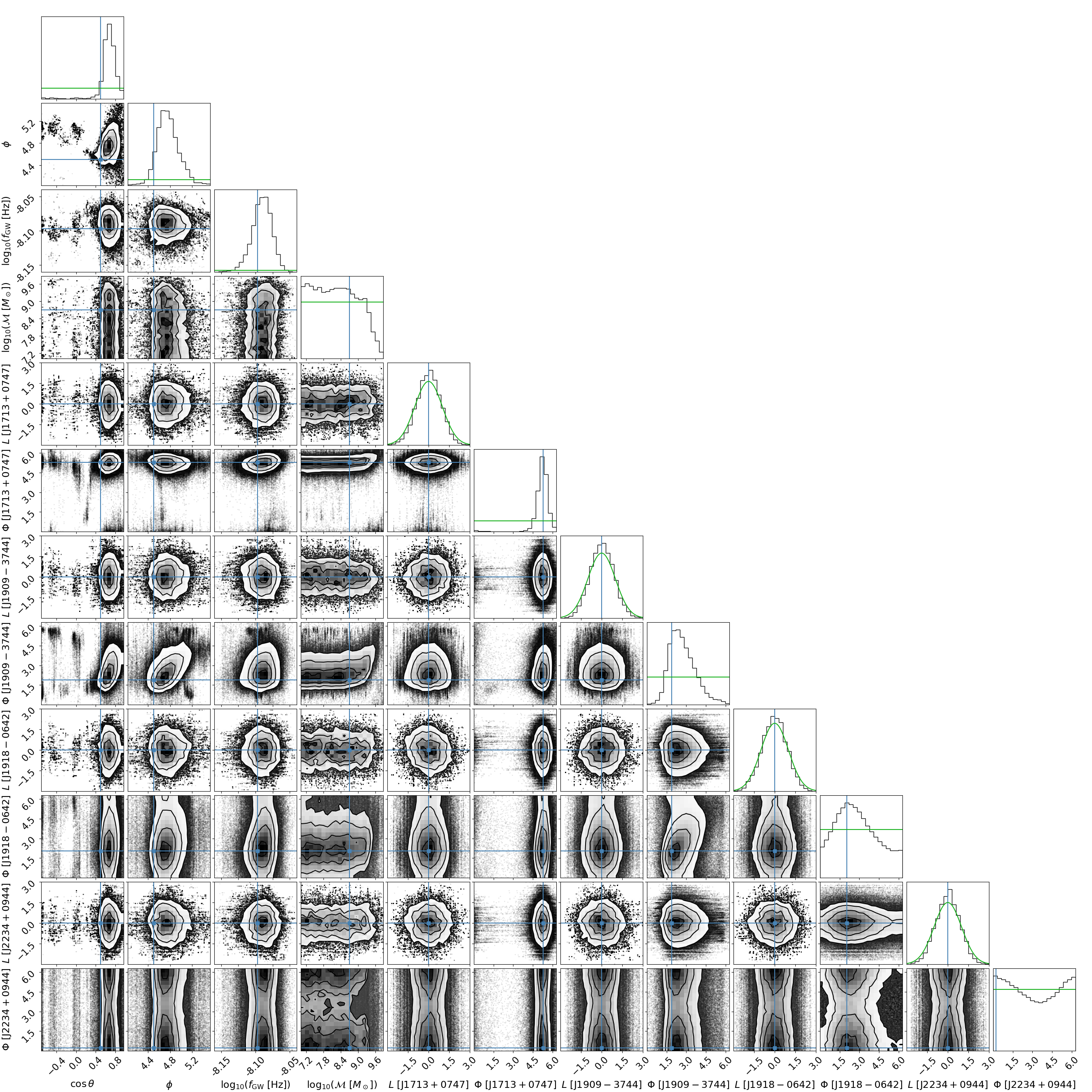}
\caption{Corner plot of four common shape parameters and selected pulsar distances and phases for the slowly evolving signal (SNR=10.4, 1 in Table \ref{tab:inj_params}). We show the posterior with black, the prior with green and the true values of parameters with blue. Contour lines represent the 1/2/3-$\sigma$ levels in two dimensions, which correspond to 39.3\%/67.5\%/86.5\% credible regions.}
\label{fig:multi_psr_corner_5e-15_slow_evolve}
\end{figure*}

\begin{figure*}[htbp]
\includegraphics[width=0.85\textwidth]{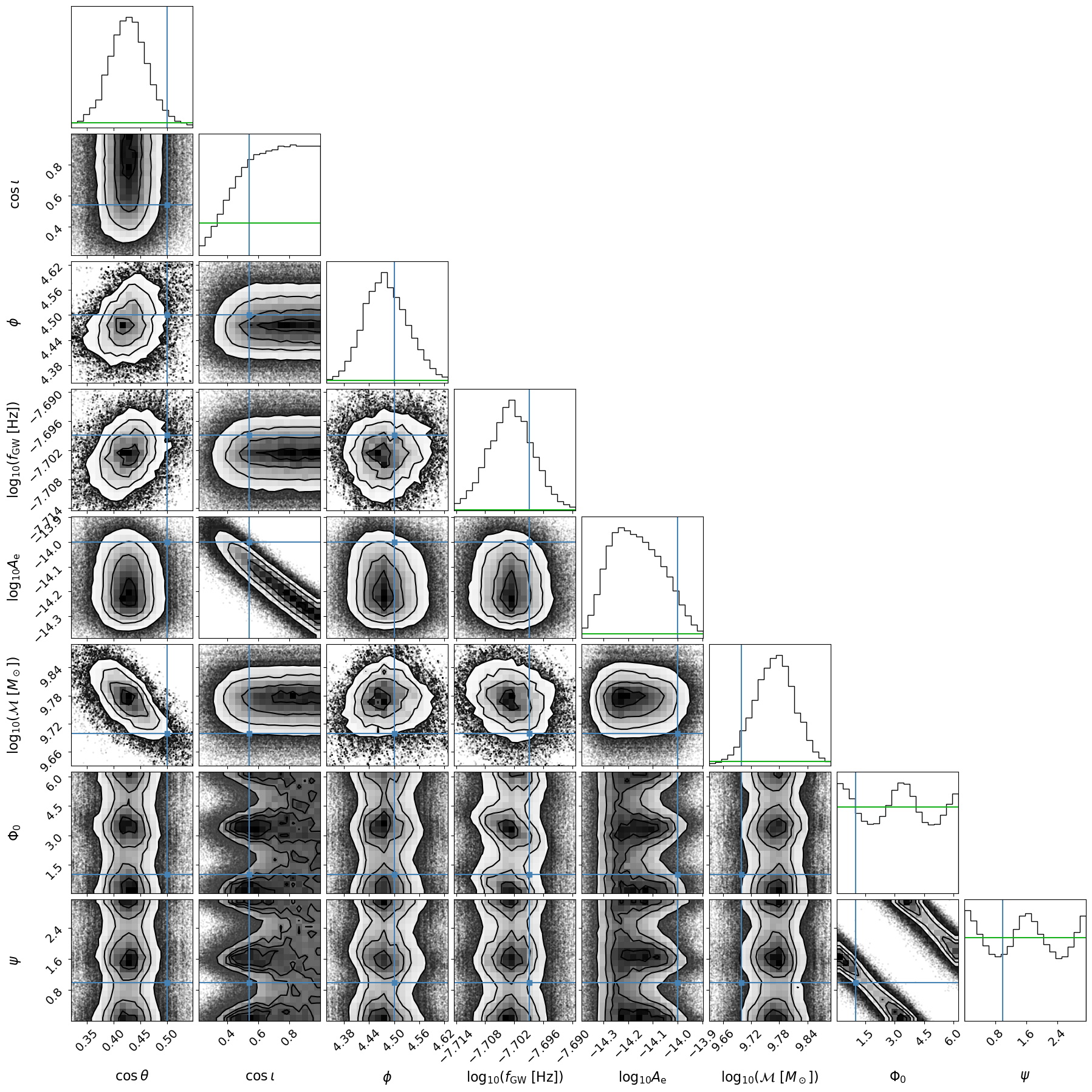}
\caption{Corner plot of parameters common to all pulsars for the fast evolving signal (SNR=13, Dataset 2 in Table \ref{tab:inj_params}). We show the posterior with black, the prior with green and the true values of parameters with blue. Contour lines represent the 1/2/3-$\sigma$ levels in two dimensions, which correspond to 39.3\%/67.5\%/86.5\% credible regions.}
\label{fig:common_par_corner_1e-14}
\end{figure*}

It is also interesting to examine the posterior distribution of some of the pulsar-specific parameters, and their correlations with some of the common parameters. Fig.~\ref{fig:multi_psr_corner_5e-15_slow_evolve} shows the pulsar distance and pulsar phase distributions for four selected pulsars, and their correlations with the four common shape parameters. We selected PSR J1713+0747 and PSR J1909-3744, as these are two of the most precise pulsars in the array and they exhibit high SNRs for this signal (4.45 and 3.25, respectively). We also show posteriors for PSR J1918-0642, which is a good example of a pulsar with a moderate SNR of 2.72, and for PSR J2234+0944 which has a negligible SNR of 0.03. We can see that accordingly, the posterior of the GW phase at PSR J1713+0747 and PSR J1909-3744 are highly peaked, while they are less informative for PSR J1918-0642 and PSR J2234+0944. Note that some of the pulsar term phases are correlated with the sky location parameters ($\theta$ and $\phi$). We can also see that at high values of $\mathcal{M}$, that also shows a correlation with the pulsar phases. These are due to the fact that we parametrize the initial phase at each pulsar as the sum of the Earth term phase, the phase collected during propagation from Earth to the pulsar, and the corresponding $\Phi_i$ parameter. Since the projected distance to the pulsar changes with sky location, the pulsar phases must be corrected as we change the sky location. Similarly, changing the chirp mass changes the phase accumulated between the Earth and the pulsar, so $\Phi_i$ needs to be adjusted.

\subsubsection{Fast evolving signal}
\label{sssec:fast_evolving}



\begin{figure*}[htbp]
\includegraphics[width=0.98\textwidth]{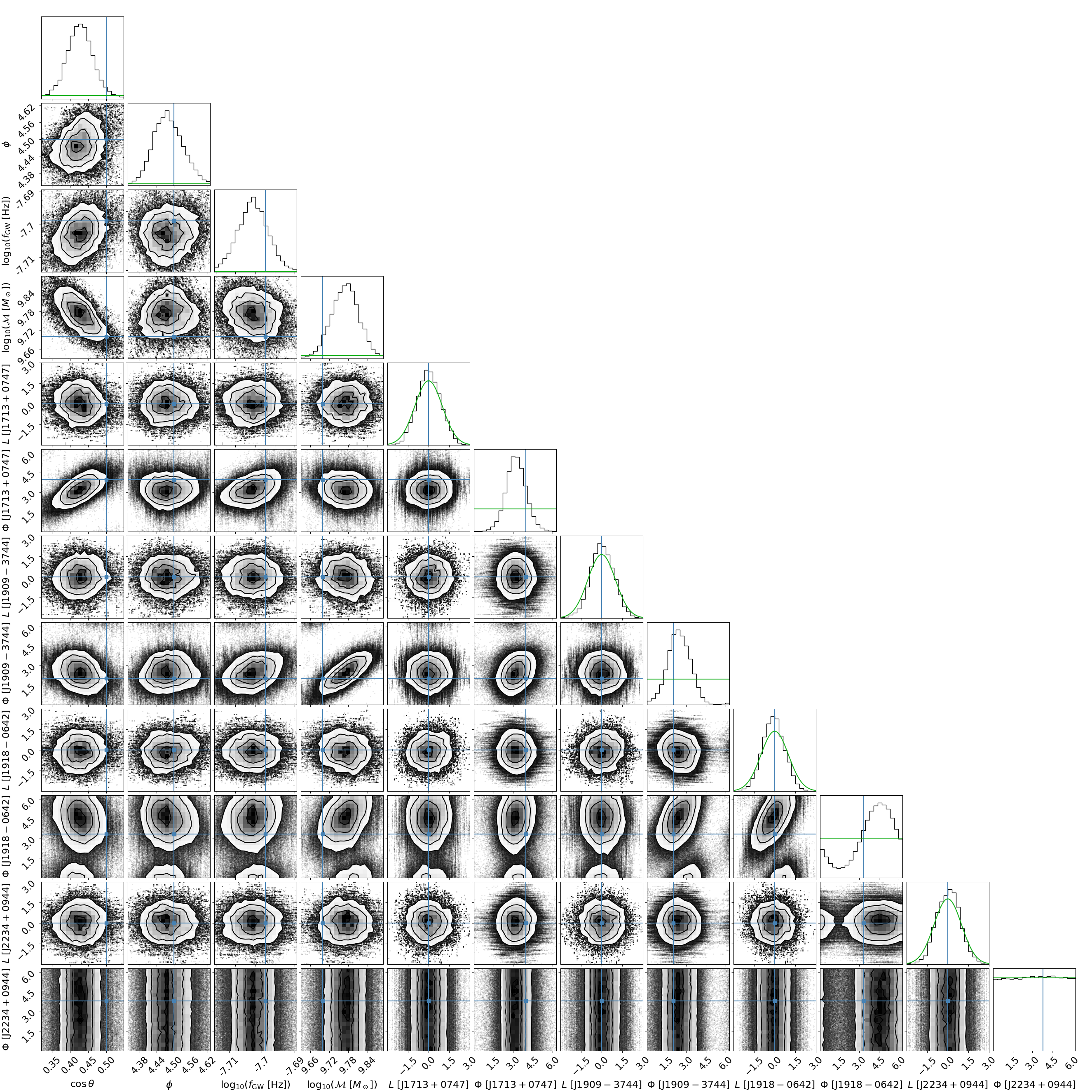}
\caption{Corner plot of four common shape parameters and selected pulsar distances and phases for the fast evolving signal (SNR=13, Dataset 2 in Table \ref{tab:inj_params}). We show the posterior with black, the prior with green and the true values of parameters with blue. Contour lines represent the 1/2/3-$\sigma$ levels in two dimensions, which correspond to 39.3\%/67.5\%/86.5\% credible regions.}
\label{fig:multi_psr_corner_1e-14} 
\end{figure*}

The next signal we analyzed has higher frequency and chirp mass than the previous one (see Dataset 2 in Table \ref{tab:inj_params}), and thus shows a significant frequency evolution between the Earth and pulsars, and it even starts to show a non-negligible evolution within the 12.5-year observing timespan (see eq.~(\ref{eq:inband_freq_evolve})). The total SNR is 13, with a highly heterogeneous distribution within pulsars. 50/90/99\% of the total SNR$^2$ comes from just 2/13/28 pulsars. We show the distribution of the eight parameters common to all pulsars in Fig.~\ref{fig:common_par_corner_1e-14}. Unlike in the slowly evolving case, the observed frequency evolution results in an informative chirp mass posterior. This moderate-SNR source results in a chirp mass measurement with a 1-$\sigma$ error of $\sim$10\%. Given the high correlations between $\mathcal{M}$ and $\cos \theta$, the chirp mass measurement precision can be further improved if the sky location of the source can be fixed. This can be done e.g.~if the host galaxy can be identified through electromagnetic observations (see e.g.~\cite{Goldstein_PTA_host_galaxy, NANOGrav_3C66B, Tingting_MMA_CW}). The intricate multimodal structure we have seen for the slowly evolving signal in Fig.~\ref{fig:common_par_corner_5e-15_slow_evolve} is much less prominent for this quickly evolving signal. Many of the one-dimensional marginal distributions we see in Fig.~\ref{fig:common_par_corner_1e-14} are close to being Gaussian, and only $\Phi_0$ and $\Psi$ show a multimodal structure. This reduction in the complexity of the posterior distributions is due to the fact that the frequency evolution breaks some of the degeneracies present in the signal model when the evolution is negligible.



\begin{figure*}[htbp]
\includegraphics[width=0.85\textwidth]{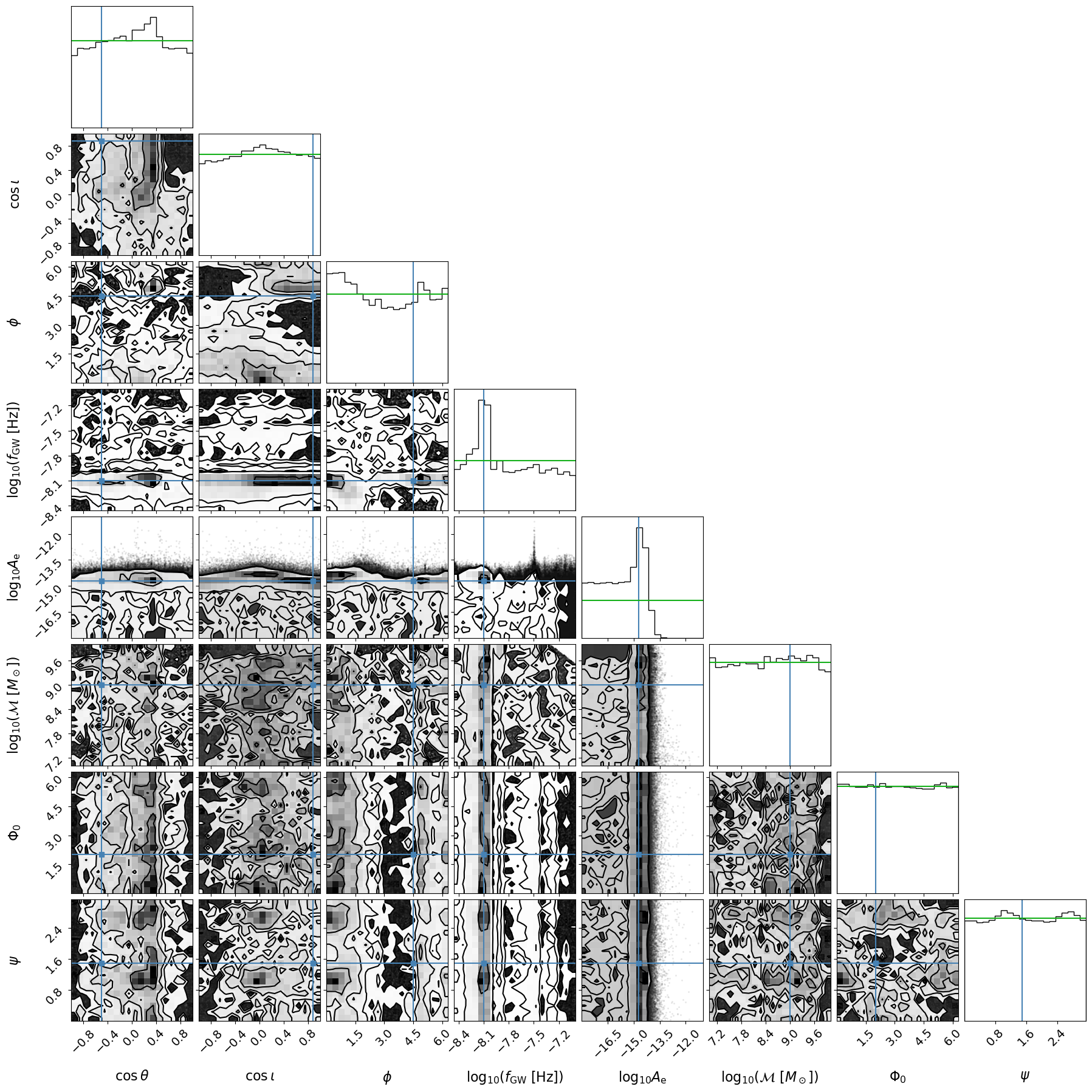}
\caption{Corner plot of parameters common to all pulsars for marginally detectable signal (SNR=4.3, Dataset 3 in Table \ref{tab:inj_params}). We show the posterior with black, the prior with green and the true values of parameters with blue.}
\label{fig:common_par_corner_2e-15_marginal}
\end{figure*}

We show the posterior distributions of phases and distances for four selected pulsars, and their correlations with the four common shape parameters in Fig.~\ref{fig:multi_psr_corner_1e-14}. We show these for the same four pulsars as in Fig.~\ref{fig:multi_psr_corner_5e-15_slow_evolve}. For this signal, they have the following SNRs: 8.3 for PSR J1713+0747, 5.42 for PSR J1909-3744, 2.17 for PSR J1918-0642, and 0.41 for PSR J2234+0944. In this example, the GW phase at PSR J2234+0944 perfectly recovers its prior. Similarly to Fig.~\ref{fig:multi_psr_corner_5e-15_slow_evolve}, we also see correlations between the pulsar phases and sky location or chirp mass. In addition, we see that pulsar phases can also be correlated with the GW frequency and pulsar distances. These have similar explanations as discussed above.

\subsubsection{Marginally detectable signal}
\label{sssec:marginal}

We also analyzed a marginally detectable signal with a total SNR of 4.3, with 50/90/99\% of the total SNR$^2$ coming from 3/12/22 pulsars (see Dataset 3 in Table \ref{tab:inj_params}). The sky location and frequency ($f_{\rm GW} = 8$ nHz) of the signal were chosen so that it roughly corresponds to the most sensitive part of the parameter space (see Section \ref{ssec:upper_limit}) and thus represents a typical signal we might expect to detect first. We have also chosen a moderate chirp mass ($\mathcal{M}=1\times10^9 \ M_{\odot}$), which is heuristically what we are most likely to detect, since higher chirp mass systems have higher GW amplitudes, but are also more rare. This combination of $f_{\rm GW}$ and $\mathcal{M}$ results in a signal with a relatively low amount of frequency evolution. 



Fig.~\ref{fig:common_par_corner_2e-15_marginal} shows the one-dimensional and 2-dimensional marginal posterior distributions of the common parameters for this signal. For most of the parameters, the posterior distributions are not significantly different from their respective priors. The main exceptions are $f_{\rm GW}$ and $A_{\rm e}$, both of which have a posterior peaked at the true location of the parameter. The posterior for $f_{\rm GW}$ also has non-negligible support over the entire prior range. $A_{\rm e}$ values significantly higher than the true parameter are ruled out, however, the posterior has support extending to the lower prior boundary, indicating the low significance of the signal. Note that the posteriors for the sky location and inclination parameters also show some deviation from the priors, but they are not particularly peaked around the true parameter values.

These analyses illustrate the expected progression of a GW detection from an individual SMBHB as we gather SNR over time. We expect to first see a peak emerging in the posteriors of $f_{\rm GW}$ and $A_{\rm e}$, while basically recovering the priors of other parameters. These other parameters will start to have more informative posteriors as we gather more SNR (as we have seen on Figs.~\ref{fig:common_par_corner_5e-15_slow_evolve} and \ref{fig:common_par_corner_1e-14}). The expected sequence of parameter constraint improvement will be important to keep in mind as we transition from placing upper limits to claiming detections. In Section \ref{ssec:upper_limit}, we also investigate how the upper limit can be affected by a marginally detectable signal.



\subsection{Upper limit analysis}
\label{ssec:upper_limit}

If no significant GW candidates are found, we can place upper limits on the amplitude of GWs from individual SMBHBs. Such an analysis poses slightly different challenges, because one needs to effectively explore the whole prior range in many parameters. Exploration of the full parameter space is required to ensure we gather a sufficient number of independent samples to get an accurate estimate of the upper limit as a function of different parameters. To test the performance of our pipeline in such a scenario, we analyzed a dataset similar to those discussed above, but with no GW signal added.

\begin{figure}[!htbp]
\includegraphics[width=\columnwidth]{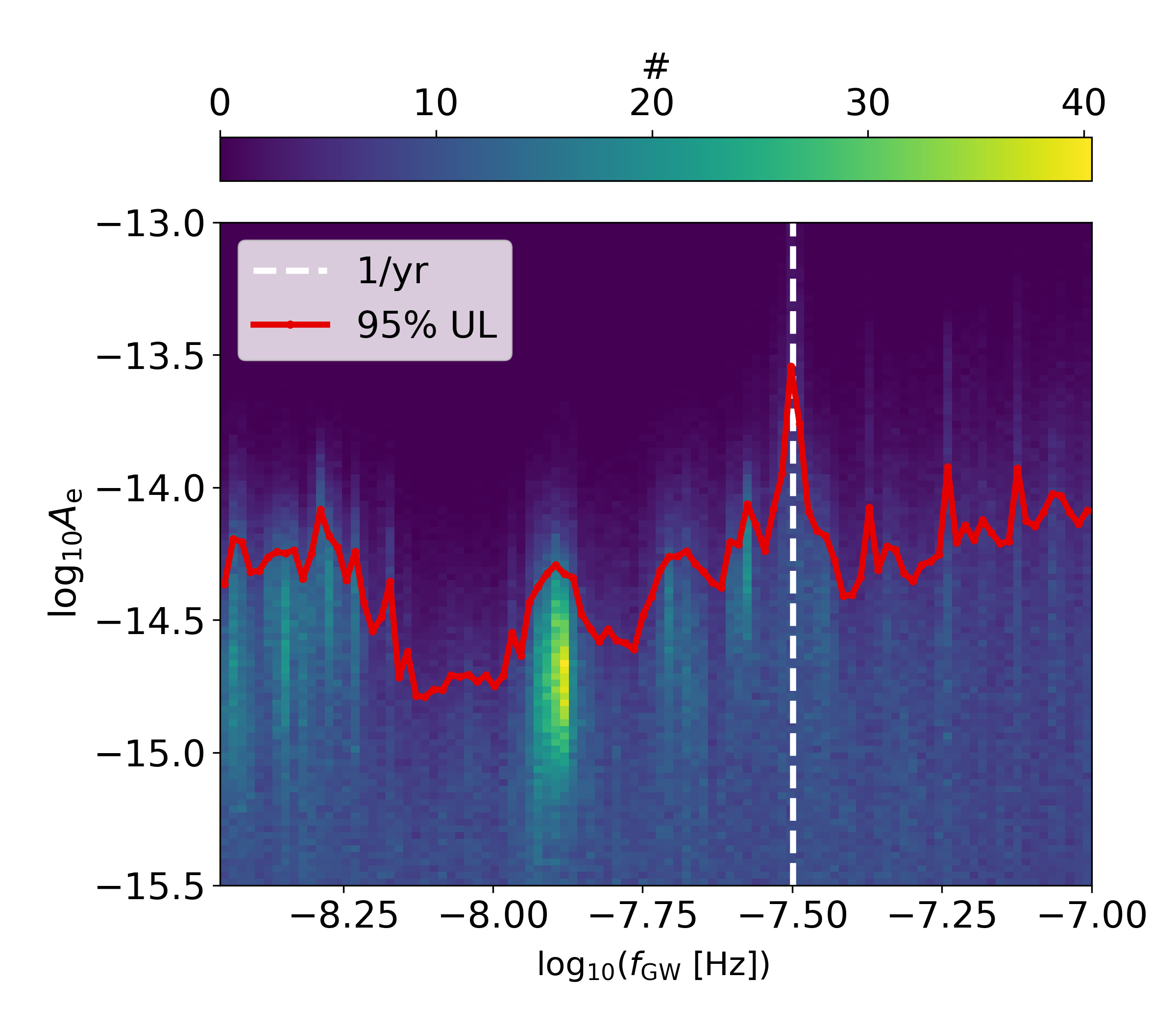}
\caption{GW amplitude upper limit as a function of GW frequency (red line). We also show the number of independent samples on the $\log_{10} f_{\rm GW}$--$\log_{10} A_{\rm e}$ plane. The vertical dashed line indicates the frequency of 1/yr, where we expect a significantly reduced sensitivity.}
\label{fig:UL_vs_freq}
\end{figure}

Fig.~\ref{fig:UL_vs_freq} shows the upper limit on $\log_{10} A_{\rm e}$ as a function of $\log_{10} f_{\rm GW}$ for this simulated dataset. We also show the number of independent samples in each bin on the $\log_{10} f_{\rm GW}$--$\log_{10} A_{\rm e}$ plane. The upper limit is calculated by binning the samples in $\log_{10} f_{\rm GW}$ and finding the 95th percentile of the amplitude in each bin. As expected, the dataset is most sensitive around 8 nHz ($\log_{10} f_{\rm GW}\simeq$ -8.1). At lower frequencies, we lose sensitivity due to the red noise present in the pulsars. At higher frequencies, we are progressively less sensitive due to the fact that we measure the integral of the GW signal. We also have particularly low sensitivity at $f_{\rm GW}=$ 1/yr due to the degeneracy with Earth's orbital period introduced when we convert the observed TOAs to the SSB. Note that there is a peak in the posterior at $(\log_{10} f_{\rm GW}, \log_{10} A_{\rm e})\simeq$ (-7.9, -14.7). The peak corresponds to a noise fluctuation being fitted with the GW model in this particular realization.

\begin{figure}[!htbp]
\includegraphics[width=\columnwidth]{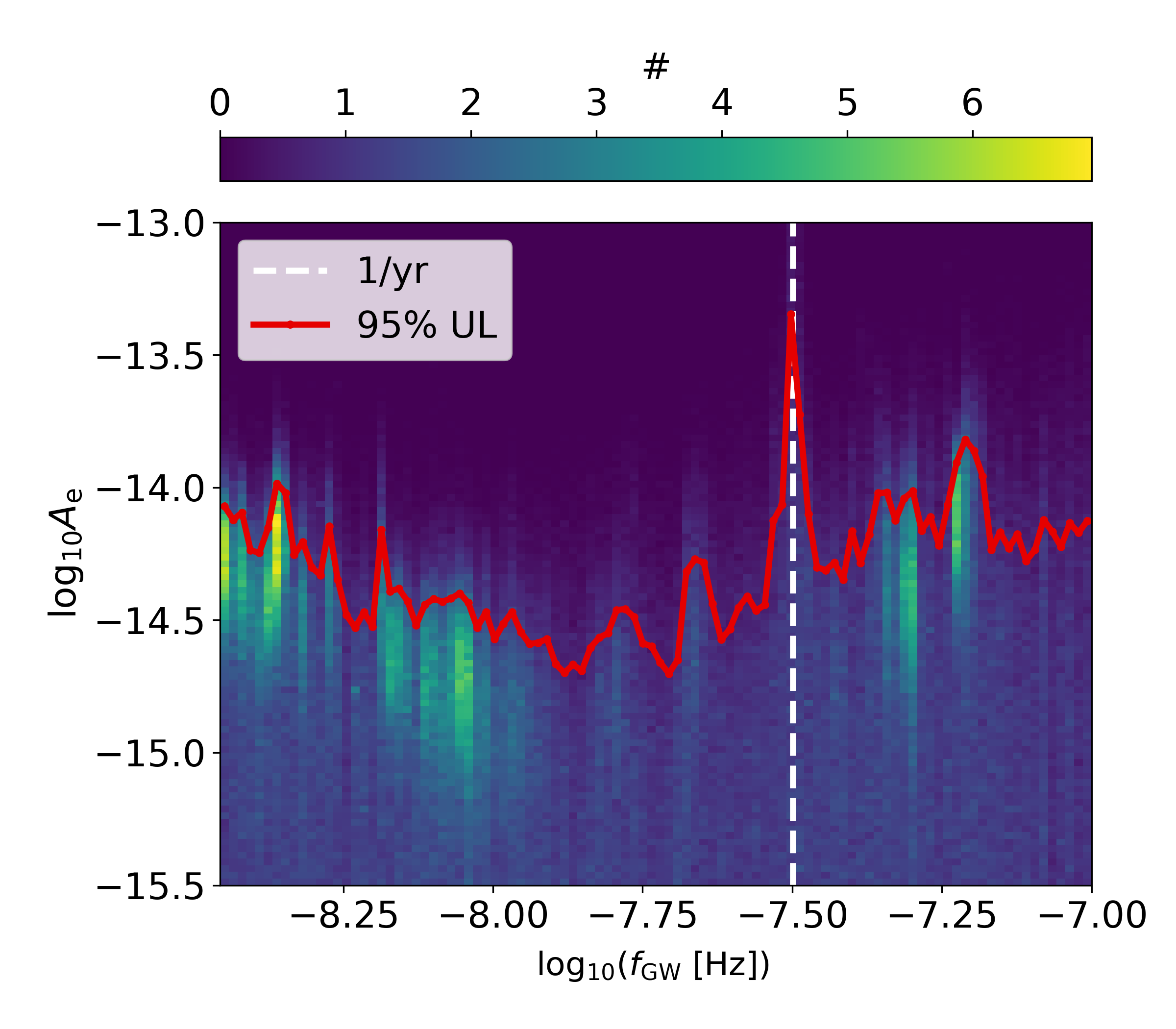}
\caption{Same as Fig.~\ref{fig:UL_vs_freq}, but with a different noise realization. Note that bumps in the upper limit due to noise fluctuations appear at different random frequencies.}
\label{fig:UL_vs_freq_new_seed}
\end{figure}

Even a pure noise dataset can exhibit features akin to marginal GWs. To investigate this feature, we analyzed a dataset with the same properties but a different random seed used for the noise realizations. Fig.~\ref{fig:UL_vs_freq_new_seed} shows the upper limit as a function of frequency for this dataset. We can see that while that peak does not show up in the alternate realization, other similar features appear at different parts of the parameter space. Candidate detections would need to be vetted with a full suite of cross-checks beyond the scope of this paper before a detection could officially be claimed. One possible approach is to reanalyze the dataset many times, while setting the sky location of the pulsars to random positions on the sky. This sky scrambling would be the same as currently used for the GWB \cite{sky_scrambles, Steve_sky_scramble}, and since it destroys the coherence of the signal, it could be used the build a null distribution. Comparing the non-scrambled result with the null distribution gives a false alarm probability of the candidate. The dramatic speedup provided by our method significantly reduces the expense of conducting such reanalyses, which will help improve the robustness of future candidate vetting.

\begin{figure}[!htbp]
\includegraphics[width=\columnwidth]{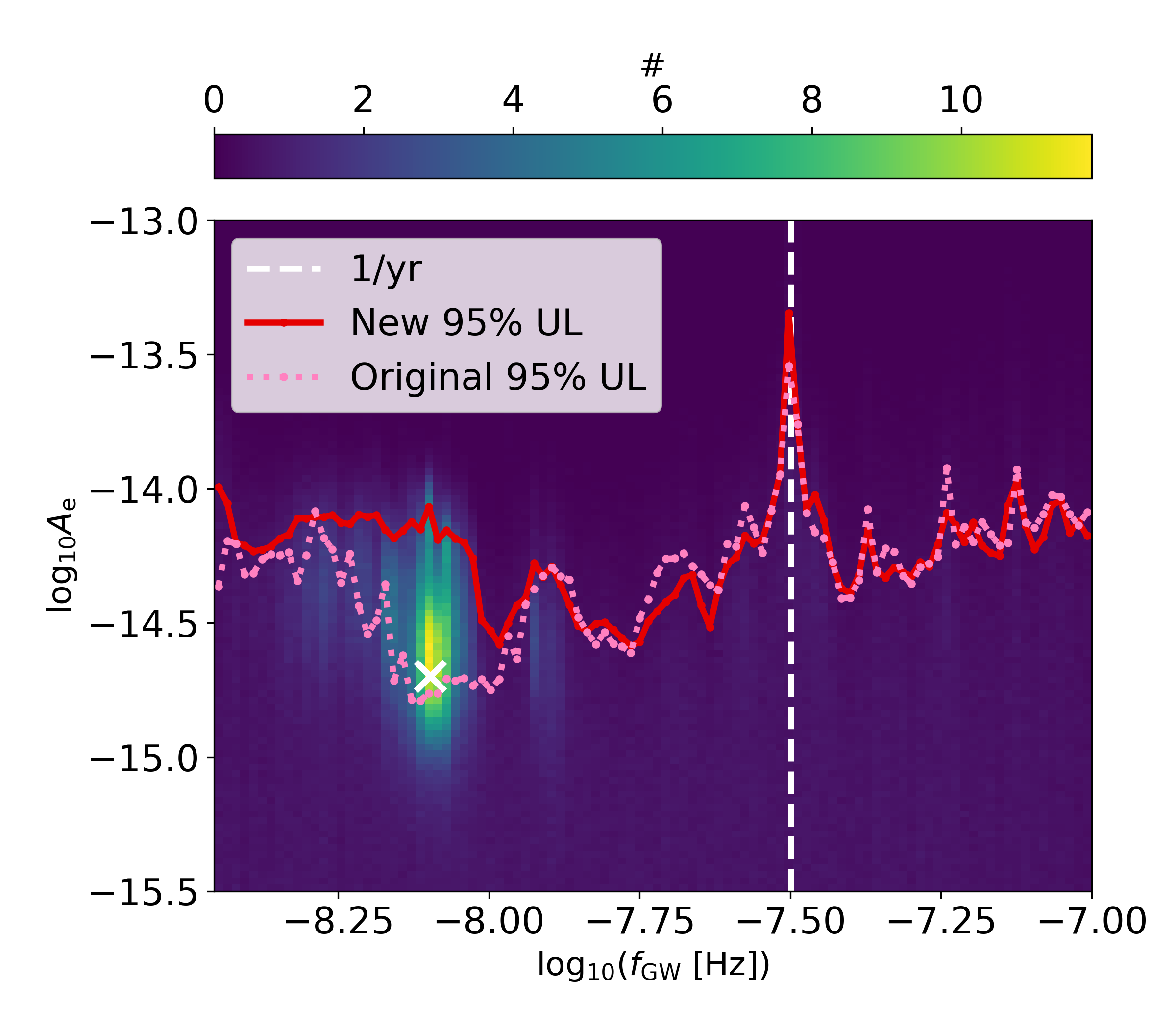}
\caption{Same as Fig.~\ref{fig:UL_vs_freq} but in the presence of a marginally detectable signal at $\log_{10} f_{\rm GW} \simeq -8.1$. The red solid line shows the upper limit we get from analyzing this dataset, which is elevated around the signal compared to the original upper limit shown by the pink dotted line. The white cross indicates the true parameters of the signal.}
\label{fig:UL_vs_freq_marginal}
\end{figure}

We also investigated how the upper limit changes in the presence of a marginal GW signal. Fig.~\ref{fig:UL_vs_freq_marginal} shows the amplitude upper limit as a function of $\log_{10} f_{\rm GW}$ for Dataset 3 from Table \ref{tab:inj_params}. As a comparison, we also show the upper limit we get from the dataset with no signal. We can see that the upper limit around the true frequency of the signal is elevated by almost an order of magnitude, even though the true amplitude is right at the level of the original upper limit. Note also that even this low-significance signal is much more prominent than the peak we have seen on Fig.~\ref{fig:UL_vs_freq} due to the noise fluctuation.

\begin{figure}[htbp]
\includegraphics[width=\columnwidth]{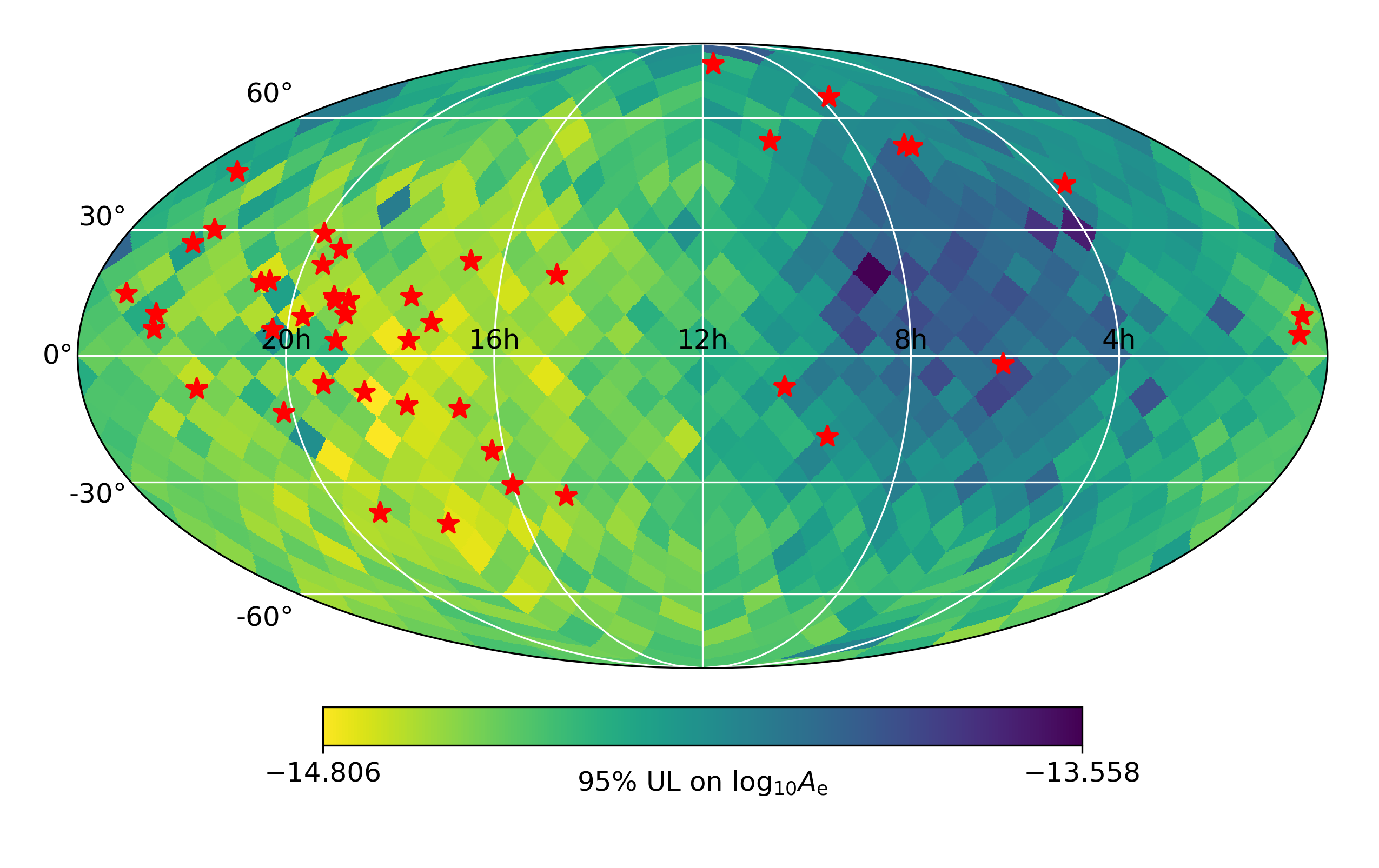}
\caption{Upper limit on $\log_{10} A_{\rm e}$  as a function of sky location. Red stars indicate locations of the pulsars in the array.}
\label{fig:UL_vs_sky}
\end{figure}

Fig.~\ref{fig:UL_vs_sky} shows the frequency-marginalized upper limit we can place on $\log_{10} A_{\rm e}$ as a function of sky location. There is more than an order of magnitude difference between the upper limit we can place towards the most and least sensitive sky location. The variable sky sensitivity is due to the fact that the pulsars in the NANOGrav 12.5-year dataset have a highly non-isotropic distribution on the sky. Such anisotropy is expected for any pulsar timing array, because there are more pulsars towards the galactic center than in the antipodal direction.

\section{Results with real data}
\label{sec:11yr_comparison}

\begin{figure}[htbp]
\includegraphics[width=\columnwidth]{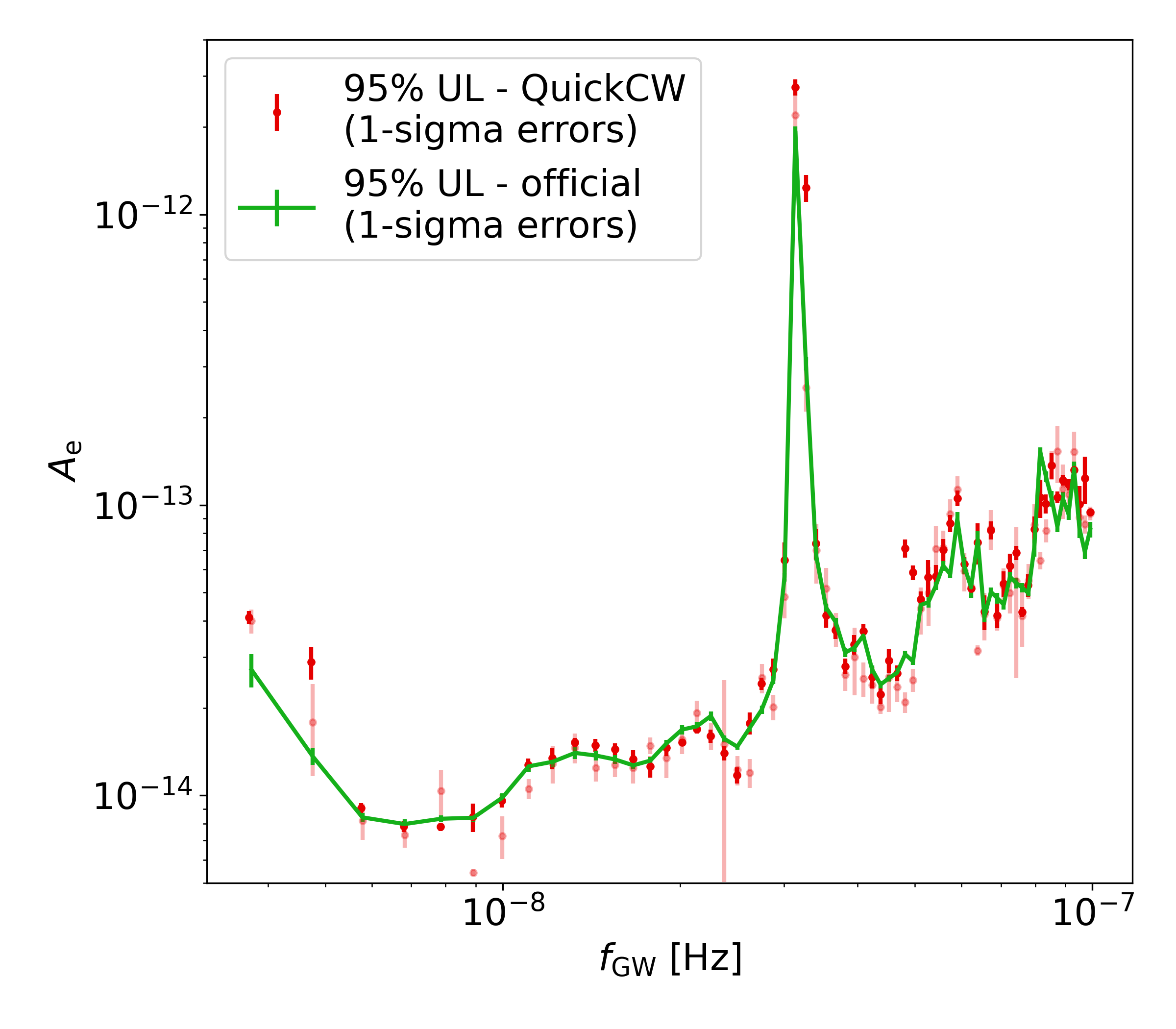}
\caption{GW amplitude upper limit as a function of GW frequency for the NANOGrav 11-year dataset analyzed by \texttt{QuickCW} (red), and as reported in Ref.~\cite{nanograv_11yr_cw} (green). We also show our results using narrow frequency bins centered at the fixed frequencies used in the official NANOGrav analysis (pink).}
\label{fig:UL_vs_freq_11yr}
\end{figure}

To validate our new methods with real data, we analyzed the NANOGrav 11-year dataset. Fig.~\ref{fig:UL_vs_freq_11yr} shows the GW amplitude upper limit we get with \texttt{QuickCW} as a function of the GW frequency in red. We also show the official NANOGrav result in green \cite{nanograv_11yr_cw}. Overall, there is good agreement between the two over the whole frequency range. Note that the official results show the amplitude upper limit at a set of fixed $f_{\rm GW}$ values, while our results show the upper limit at frequency bins with non-negligible width. Thus we do not expect perfect agreement between the two results, especially where the upper limit is quickly changing with frequency. To illustrate this, we also plot our results using narrow frequency bins around the fixed frequencies used in the official NANOGrav analysis (pink markers). We can see that at several frequencies, these are in better agreement with the official results (e.g.~around $f_{\rm GW}=1$/yr). However, even these narrow frequency bin results show some discrepancy at the lowest frequency bin. We think this is due to a bug that was recently found in the so called empirical distribution proposals, which were used in the official analysis \cite{nanograv_11yr_cw}\footnote{Note that this issue was recently resolved here: \url{https://github.com/nanograv/enterprise_extensions/commit/3a17943dddfa005867cdbaa960d4e72b747eb373}}. The bug resulted in a small overestimation of the red noise in some pulsars, which in turn meant that the upper limit on $A_{\rm e}$ was underestimated at low frequencies, where there is a strong correlation between the red noise and the individual binary models.

Allowing the GW frequency to explore the entire prior range has several advantages over running at a set of fixed frequencies. Especially at frequencies where the upper limit changes quickly, a fixed frequency analysis cannot fully explore parameter space and artificially underestimates the uncertainty in the upper limits. Varying the frequency also streamlines the analysis, since we only need a single run instead of dozens. A potential issue is that we do not get the same number of samples at all frequencies. Excessive focus on a particular set of frequencies is sub-optimal, since if we require a fixed level of convrgence in each frequency bin, the overall runtime is determined by the frequency bin with the least number of samples. To optimize the sampling, we could apply a pseudo-prior on $f_{\rm GW}$, which down-weights regions with many samples based on a pilot run. The pseudo-prior would then ensure a more uniform number of samples at all frequencies while preserving good mixing. The undesired bias from the presude-prior can be canceled in post processing by reweighting the posterior samples. This technique is similar to umbrella sampling \cite{umbrella_sampling}, and the pseudo-prior used when calculating Bayes factors with the product space method \cite{product_space}.

\section{Conclusion}
\label{sec:conclusion}

In this paper, we presented a new formulation of the likelihood function (see eq.~(\ref{eq:logL})), which results in a significant speedup of a Bayesian search for individual SMBHBs in PTA data. Our formulation does not apply any new approximations, and recovers the canonical likelihood within numerical errors. This is achieved by separating the parameter space to shape parameters and projection parameters. Precalculating inner products for a given set of shape parameters allows cheap exploration of the projection parameters, thus speeding up the entire analysis. We demonstrated the performance using a new analysis pipeline employing the new likelihood in a Metropolis-within-Gibbs sampler with multiple-try MCMC.

These methods will drastically reduce the computational cost of searches for individual sources in upcoming PTA datasets, and improve the tractability of achieving a well-converged analysis as the size of datasets increases. Since all the latest PTA datasets show evidence for the presence of a common red noise process \cite{nanograv_12p5yr_gwb,epta_dr2_gwb, ppta_dr2_gwb, ipta_dr2_gwb}, we plan to incorporate that in our model as well. This would appear as an additional red noise term in the $T$ and $B$ matrices defined below eq.~(\ref{inner_product}), and would introduce two additional parameters to sample over, which describe the amplitude and spectral slope of the common red noise. This addition would not significantly change how the fast likelihood methods presented in this paper work. If the upcoming PTA datasets will show evidence for this common process being correlated between pulsars, we would ultimately want to include those correlations in our model as well. In Appendix \ref{sec:gwb_appendix} we outline how such correlations could be incorporated into the fast likelihood framework we presented in this paper. The implementation of that addition will be presented in future work. 

These methods will also make more sophisticated analyses computationally feasible. In particular, we plan to extend these methods to work with \texttt{BayesHopper} \cite{BayesHopper}, a pipeline proposed to search for multiple individual sources simultaneously. We also plan to work on implementing similar methods for a search for eccentric SMBHBs and a search for non-Einsteinian polarization modes from SMBHBs \cite{Logan_CWAltPol}. It might also be worthwhile to extend these methods to sine-Gaussian wavelets in an effort to speedup \texttt{BayesHopperBurst} \cite{BayesHopperBurst}, a pipeline to search for generic GW bursts in PTA data.

\begin{acknowledgments}
The authors thank the organizers of the 2021 NANOGrav Fall Meeting, which provided inspiration for this project. We thank Ken Olum and Stephen Taylor for feedback on the manuscript, and Sarah Vigeland for helping reproduce the NANOGrav 11-year results. We used the \texttt{corner} \cite{corner} software package to produce the corner plots presented in this paper. Some of the results in this paper have been derived using the \texttt{healpy} \cite{healpy} and \texttt{HEALPix} \cite{healpix} package. We appreciate the support of the NSF Physics Frontiers Center Award PFC-1430284 and NASA LISA foundation Science Grant 80NSSC19K0320.
\end{acknowledgments}

\appendix

\section{Fast likelihood in the presence of correlated red noise}
\label{sec:gwb_appendix}

A red noise process can be modeled using a pseudo Fourier basis ${\bf F}$, made up of a collection of sines and cosines at a discrete set of frequencies with amplitudes ${\bf a}$~\cite{Lentati:2012xb}. The log likelihood then becomes:
\begin{equation}
\log L = -\frac{1}{2} (\delta t - s - {\bf F}\cdot{\bf a} |\delta t -s - {\bf F}\cdot{\bf a}) \, .
\label{eq:F-basis}
\end{equation}
The usual approach is to marginalize over the ${\bf a}$'s analytically. If we assume the signal is not correlated between pulsars, eq.~(\ref{eq:F-basis}) can still be evaluated pulsar-by-pulsar \cite{FactorizedLikelihood}.  However, the per-pulsar factorization breaks down if we intend to model a GWB with Hellings-Downs correlations between the pulsars \cite{Cornish:2013nma}. This would significantly increase the cost of the CW analysis since the $M^{kl}$ matrix would become dense.

Another approach is to use a method similar to what we are proposing for individual SMBHBs and to the methods presented in Ref.~\cite{CorrelatedFactorizedLikelihood}. For simplicity let us ignore the individual sources for now and focus on the GWB. Expanding the log likelihood we have:
\begin{equation}
\log L= -\frac{1}{2} \left[ (\delta t|\delta t)  -2{\bf a} \cdot {\bf P} + {\bf a} \cdot {\bf Q} \cdot {\bf a}  \right] \, ,
\end{equation}
where ${\bf P}  = (\delta t | {\bf F})$ and ${\bf Q}  = ({\bf F} | {\bf F})$.
Note that the vector ${\bf P}$ is much smaller than the vector ${\bf F}$. The vector ${\bf F}$ is a vector of vectors, made up of the time samples for the sines and cosines at each frequency and for each pulsar. After the inner products have been done, the ${\bf P}$ is a collection of numbers, one each for the sine and cosine at each frequency in each pulsar. The matrix ${\bf Q}$ has rows and columns that follow the pattern of the vector ${\bf P}$. To simplify the discussion, consider a single frequency $f$, and label the ${\bf a}$ such that $a_{(2i-1)}$ are the cosine terms in pulsar $i$ and $a_{(2i)}$ are the sine terms. Our hyper-prior for the ${\bf a}$ is such that each $a_k$ is drawn from a Gaussian distribution with variance $S(f)$ and correlations given by:
\begin{eqnarray}\label{corr}
{\rm E}[a_{(2i-1)} a_{(2j - 1)} ] & = & H_{ij} S(f) \nonumber \\
{\rm E}[a_{(2i-1)} a_{(2j )} ] & = & 0 \nonumber \\
{\rm E}[a_{(2i)} a_{(2j)} ] & = & H_{ij} S(f) \, ,
\end{eqnarray}
where $H_{ij}$ is the Hellings-Downs correlation between pulsars $i,j$. More schematically we can write ${\rm E}[a_k a_l ] = \varphi_{kl}$ where the entries of $\varphi_{kl}$ are given by (\ref{corr}). The posterior probability distribution for ${\bf a}$ can then be written as:
\begin{eqnarray}
&&p({\bf a}| \delta t, S) = \frac{e^{-\frac{1}{2}(\delta t | \delta t)}}{\sqrt{ {\rm det}(2\pi C)  {\rm det}(2\pi \varphi)}} \nonumber \\
 && \hspace{1in} \times e^{a_k P^k - \frac{1}{2} a_k a_l  Q^{kl}-\frac{1}{2} a_k a_l (\varphi^{-1})^{kl}} \, .
\end{eqnarray}
The next step is to marginalize the $p({\bf a}| \delta t, S)$ over ${\bf a}$ to produce the marginal likelihood $p(\delta t | S)$. This marginalization can be done analytically since the expression for $p({\bf a}| \delta t, S)$ is a multi-variate Gaussian. The end product is the new likelihood for the power at frequency $f$. Notice that all the inner products were done pulsar by pulsar. We have a factorized likelihood. Unlike in the usual analysis where the marginalization is done before computing the inner products, by performing the marginalization after the per-pulsar inner products have been done we avoid any cross terms. The analysis still ``knows'' about the correlations, without actually having to cross-correlate the data between pulsars.

Note that the inner products ${\bf P}$ and ${\bf Q}$ have to be recomputed every time the noise model is updated. The red noise can be treated like the GWB, but just with a diagonal correlation matrix, so ${\bf P}$ and ${\bf Q}$ only have to be recalculated for white noise updates. The white noise inner products could be computed for some discrete set of values for each pulsar, then a joint marginalization across the white noise in all the pulsars could be done using a lookup table.

\begin{table*}[htbp]
\caption{\label{tab:runtimes}%
Runtimes of old likelihood and different components of the new likelihood for various representative datasets\label{tab:timings}
}
\begin{ruledtabular}
\begin{tabular}{c|ccc}
&
\textrm{NANOGrav 12.5-year}&
\textrm{2 $\times$ NANOGrav 12.5-year}&
\textrm{5 $\times$ NANOGrav 12.5-year}\\
\colrule
\# of pulsars & 45 & 90 & 225\\
\# of TOAs & 410,064 & 820,128 & 2,050,320\\
\colrule
Old likelihood GW update & 300 ms & 610 ms & 1500 ms\\
Old likelihood red noise update & 120 ms & 210 ms & 630 ms\\
\colrule
New likelihood\footnote{With fixed shape parameters} & 0.016 ms & 0.020 ms & 0.029 ms\\
GW shape parameter update\footnote{Common parameters or pulsars distances} & 90 ms & 220 ms & 770 ms\\
Pulsar red noise update & 180 ms & 400 ms & 1300 ms\\
\end{tabular}
\end{ruledtabular}
\end{table*}

\section{Implementation in \texttt{Python} using \texttt{Numba}}
\label{sec:implementation_appendix}

The new formulation of the likelihood and the associated sampler is implemented in the \texttt{QuickCW} package. \texttt{QuickCW} relies on the \texttt{ENTERPRISE} software \cite{enterprise}, which uses \texttt{Python}, so it was a natural choice to implement \texttt{QuickCW} in \texttt{Python} as well. To overcome the inherent speed limitations of \texttt{Python}, we use the \texttt{Numba} software package \cite{numba_paper, numba_zenodo}, which is a just-in-time compiler capable of generating fast machine code from \texttt{Python} syntax. This is particularly important, as in the new formulation of the likelihood, the computational cost after the inner products have been precomputed consists of a small number of simple multiplicative and additive operations. As a result, run times are dominated by \texttt{Python}-specific overheads if we do not use \texttt{Numba}.

Table \ref{tab:timings} shows representative runtimes of the old and new likelihood for different scenarios. We carried out these tests on an AMD Ryzen Threadripper 3970X 32-core processor. We tested on three different simulated datasets: one made to resemble the NANOGrav 12.5-year dataset (same as used for all other results in this paper);  one with double the number of pulsars and TOAs as the NANOGrav 12.5-year dataset; and one with five times as many pulsars and TOAs as the NANOGrav 12.5-year dataset. The latter two were used to show how performance will change in the future as the PTA datasets grow in size.

For each of these datasets we timed the execution of the old likelihood for randomly drawn GW parameters and randomly drawn red noise parameters. We can see in Table \ref{tab:timings} that these take a different amount of time. The variation is due to the fact that \texttt{ENTERPRISE} caches some of its internal function calls, so when the red noise is changed, some parts of the likelihood can be reused from before, which results in different execution times. We also tested the new likelihood formulation in a scenario when only projection parameters are changed. We can see that for the dataset resembling the NANOGrav 12.5-year, this is $\sim$20,000 times faster than the old likelihood. We also timed the recalculation of the inner products, which is necessary when the shape parameters are updated. We can see that recalculating the filters for a new set of shape parameters is still about a factor of three faster than the old likelihood. The speedup is due to a combination of algorithmic optimization and the speedup we get with \texttt{Numba}. If GW parameters are updated, the $N^k$ and $M^{kl}$ needs to be recomputed using the updated $S^k$, but one can use $C^{-1}$ from memory. However, when the red noise parameters are updated, $S^k$ remains the same, but an updated $C^{-1}$ needs to be used. As we can see in Table \ref{tab:timings}, this results in the red noise updates being more expensive.

The results for larger datasets show that runtimes for the old likelihood roughly scale linearly with the number of pulsars and TOAs. Evaluating the new likelihood at given shape parameters scales slower than linear, resulting in less than twice as long runtimes for the 5 times larger dataset. On the other hand, shape parameter updates scale faster than linear, resulting in about a factor of 8 slower evaluation for the 5 times larger dataset. Note, however, that we optimized the pipeline for a dataset like the NANOGrav 12.5-year, so the performance for larger datasets might be sub-optimal.


\end{document}